\documentclass[10pt, conference]{IEEEtran}  
\IEEEoverridecommandlockouts
\usepackage{acronym}
\usepackage{tikz}
\usetikzlibrary{shapes,snakes}
\usepackage{empheq}
\usepackage{graphicx}
\usepackage{epstopdf}
\usepackage{multirow}
\usepackage{tabu}
\usepackage{array}
\usepackage{url}
\usepackage{amsmath}
\usepackage{amssymb}
\usepackage{amsthm}
\usepackage{amsfonts}
\usepackage{tikz}
\usepackage{bm}
\usepackage{algorithmic}
\usepackage[ruled,vlined,linesnumbered]{algorithm2e}
\usepackage{graphicx}
\usetikzlibrary{arrows}
\usepackage{cite}
\usepackage{caption}
\usepackage{color}
\usepackage{amssymb}
\usepackage{tabulary}
\usepackage{booktabs}

\usepackage[justification=centering]{caption} 
\DeclareCaptionLabelFormat{algnonumber}{algorithm}
\tikzstyle{int}=[draw, fill=white!20, minimum size=2em]
\tikzstyle{init} = [pin edge={to-,thin,black}]

\newcommand\blfootnote[1]{%
	\begingroup
	\renewcommand\thefootnote{}\footnote{#1}%
	\addtocounter{footnote}{-1}%
	\endgroup
}

\newif\iftodo   
\todotrue
\newif\iftodoshort  
\todoshortfalse
\newcommand{\ma}[1]{\boldsymbol{\mathbf{#1}}}
\newcommand{\ve}[1]{\boldsymbol{\mathbf{#1}}}

\newcommand{\set}[1]{\mathcal{#1}}

\usepackage{subfig}

\begin{document}
	\title{
 \LARGE	Traversing Virtual Network Functions from the Edge to the Core: \\ \noindent
 An End-to-End  Performance Analysis
}

\author{
	\IEEEauthorblockN{Emmanouil Fountoulakis\IEEEauthorrefmark{1},
		Qi Liao\IEEEauthorrefmark{2},
		Manuel Stein\IEEEauthorrefmark{2},
		Nikolaos Pappas\IEEEauthorrefmark{1}
		\IEEEauthorblockA{\IEEEauthorrefmark{1}Department of Science and Technology, Link{\"o}ping University, Sweden}
		\IEEEauthorblockA{\IEEEauthorrefmark{2}Nokia Bell Labs, Stuttgart, Germany}
		E-mails: \{emmanouil.fountoulakis, nikolaos.pappas\}@liu.se,
		 \{qi.liao, manuel.stein\}@nokia-bell-labs.com}
}
\maketitle
	
\begin{abstract}
  Future mobile networks supporting Internet of Things are expected to provide both high throughput and low latency to user-specific services. One way to overcome this challenge is to adopt network function virtualization and Multi-access Edge Computing (MEC). In this paper, we analyze an end-to-end communications system that consists of both MEC servers and a server at the core network hosting different types of virtual network functions. We develop a queueing model for the performance analysis of the system consisting of both processing and transmission flows. We provide analytical approximations of the performance metrics such as system drop rate and average number of tasks in the system. Simulation results show that our approximations perform quite well. By evaluating the system under different scenarios, we provide insights for the decision making on traffic flow control and its impact on critical performance metrics.
\end{abstract}
	 
	
	\IEEEpeerreviewmaketitle 
	
\section{Introduction}
\vspace{0mm}
\blfootnote{
This work has been supported by the European Union’s Horizon 2020 research and innovation programme under the Marie Sk\l{}odowska-Curie grant agreement No. 643002.}
In future communications systems, mission-critical mobile applications, e.g., augmented reality, connected vehicles, eHealth, will provide services that require ultra-low latency \cite{MECKeyTech}, \cite{taleb2017multi}. To satisfy the low latency requirements, \ac{MEC} has been proposed as a key solution \cite{MECKeyTech}. The idea of MEC is to locate more computational resources closer to the users, e.g., at the base stations. Besides  latency constraints, these services may have strict function chaining requirements. In other words, each service has to be processed by a set of network functions (e.g., firewalls, transcoders, load balancers, etc.) in a specific order \cite{NFV_StateOfTheArt}. 
Furthermore, the requirements of 5G networks for flexibility and elasticity of the network inspire the idea of \ac{NFV} \cite{NFV_StateOfTheArt}, \cite{IntegratedNFV_SDN}. The idea of NFV is to decouple the network functions from dedicated hardware equipment. Instead of dedicated hardware equipment, general purpose servers can host one or more types of network functions. However, the computational	 capabilities and the available resources of MEC servers are still limited compared to the high-end servers in the cloud. Therefore, it is interesting to further investigate the cooperation between the edge and the core, and the cooperation among MEC servers.

Recently, \ac{VNF}  placement and resource allocation problem has attracted a lot of attention, e.g., \cite{NearOptPlac, JointOptJournal}. In these works, the authors formulate the VNF placement problem as mixed integer linear problem  under the assumption of known traffic demand.
In a  dynamic environment with unknown traffic, the authors in \cite{feng2018optimal} develop dynamic algorithms in order to control the flow by applying Lyapunov optimization theory.
There are few works on analyzing networks and deriving key performance metrics such as delay. Authors in \cite{YeE2DDelay} analyze the end-to-end delay for embedded VNF chains. They consider two types of services that traverse different VNF chains and provide the delay analysis for each chain. However, this work considers a specific system model where multiple VNF chains embedded on a common determined network path, while routing and flow control are not considered. Furthermore, the authors in  \cite{ApproximatingBegin} and \cite{EvaluationBegin} estimate the end-to-end delay in \ac{SDN} environment by using local node measurements for single flow and multi-flow cases, respectively.

In this paper, we model and analyze an  end-to-end communications system  consisting of two MEC servers at the edge network and one at the core network hosting different types of VNFs by applying tools from queueing theory. In order to simplify the analysis, we introduce the approach of decomposing the system into subsystems,  which can be further applied in analyzing scale-up system with arbitrary number of servers. We provide analytical expressions for the key performance metrics such as 	average number of tasks in the system and system drop rate for each subsystem. Simulation results validate our analysis and show that our analytical model is accurate. Furthermore, by evaluating the system under different scenarios we provide insights of how the routing decision affects the key performance metrics of our interest.


\section{System Model}

\begin{figure}[t!]
	\includegraphics[scale=0.36]{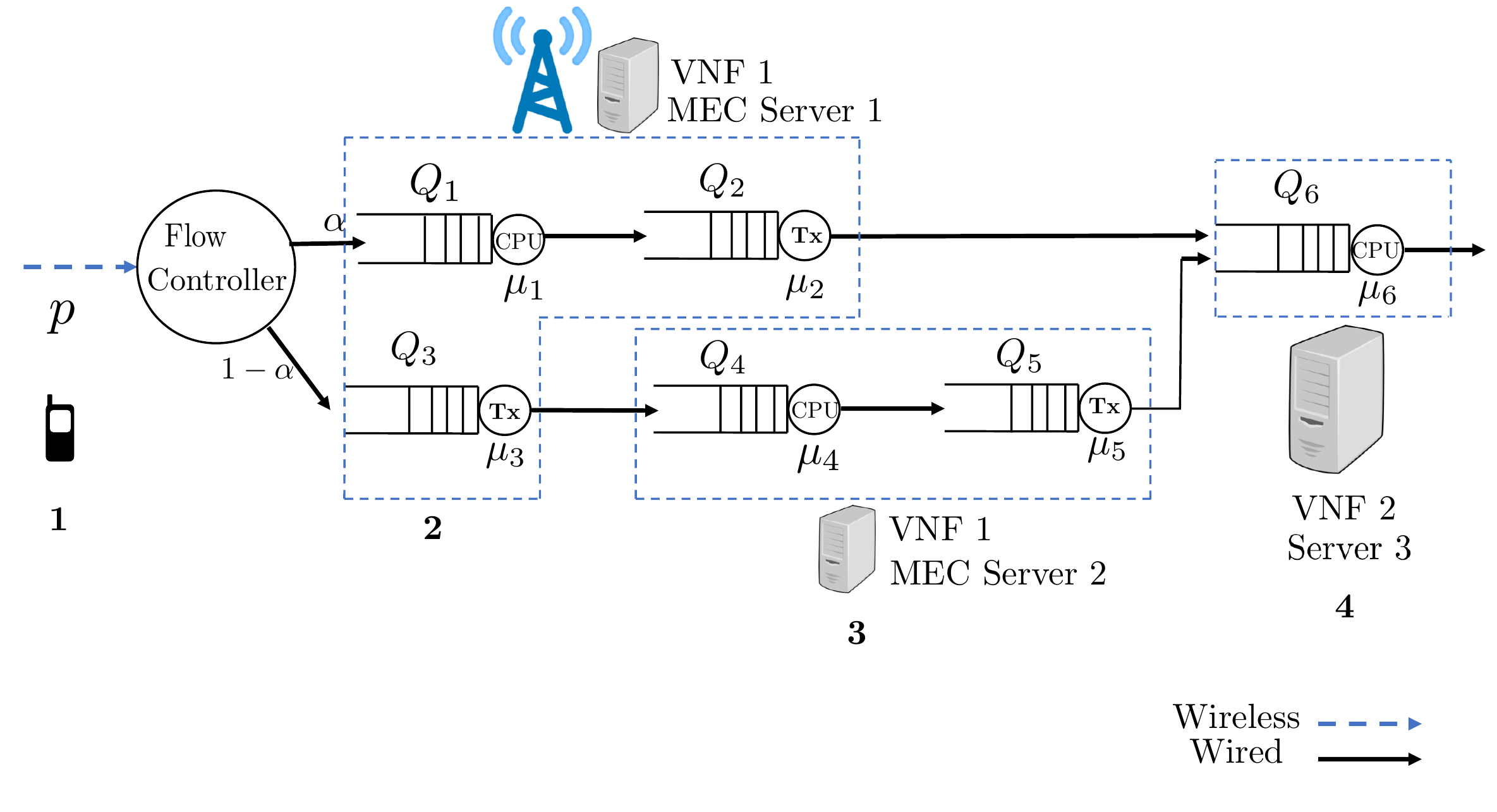} 
	\caption{The system model. The blue dashed lines group the queues located in the same server.}
	\label{Fig: System}
\end{figure}

 
We consider an  end-to-end communications system consisting of a mobile device, two \ac{MEC} servers, and one server located in the core network as depicted in Fig. \ref{Fig: System}.
A task traverses a service chain of two consecutive VNFs: \ac{VNF} $1$ and \ac{VNF} $2$. In this system, an \ac{MEC} server, called Server $1$, is co-located with the base station and hosts one copy of VNF $1$ as the primary MEC server. A secondary \ac{MEC} server, called Server $2$, is located nearby and also hosts a copy of \ac{VNF} $1$. In addition, Server $3$ in the core network hosts VNF $2$ and has more advanced computational capabilities than Servers $1$ and $2$.
We assume a slotted time system. At each time slot, the device transmits a task in form of a packet to a base station over a wireless channel. Because of the presence of fading in the wireless channel, transmissions may face errors. A task is successfully transmitted to the base\vspace{0mm} station with a probability $p$ that captures fading, attenuation, noise, etc. The device attempts for a new task transmission only if the previous task is successfully received at the base station.
The received  tasks need to be distributed between the queue for local processing and the queue for transmission to the secondary MEC server.
Thus, there are two possible routes to pass through the service chain.
A flow controller at the base station decides randomly the routing for each task\footnote{Probabilistic routing is a common strategy widely used in the literature, see for example \cite{ploumidis2016flow}.}. With probability $\alpha$ the task is processed by Server $1$ first, and then forwarded to Server $3$. With probability $1-\alpha$ the task is forwarded to Server $2$, to be processed by VNF $1$, and then forwarded to Server $3$ for being processed by VNF $2$.   

Each task that arrives at a server first waits in a queue for being processed by a \ac{VNF}. Then, after the processing, it is stored in the transmission queue, waiting to be forwarded and processed by the next \ac{VNF}. 
Let $Q_{i}$ denote the $i$-th queue, where $i \in \mathcal{K}$, and $\mathcal{K}$ is the set of the queues in the system.
Note that the queues follow an early departure-late arrival model: at the beginning of the slot the departure takes place and a new arrival can enter the queue at the end of the slot. 
The queues for task transmission are $Q_{2}$, $Q_{3}$, and $Q_{5}$, and the queues for task processing are $Q_{1}$, $Q_{4}$, $Q_{6}$. 
The arrival rates for queues $Q_1$ and $Q_3$ are $p\alpha$ and $p(1-\alpha)$, respectively. We denote by $\mu_{i}$, $i \in\set{K}$, the service rates of the queues. We assume that the service times are geometrically distributed. Furthermore, given that $Q_1$, $Q_3$, and $Q_4$ are non empty, the arrival rates of $Q_2$, $Q_4$, and $Q_5$ are equivalent to the service rates of $Q_1$, $Q_3$, and $Q_4$ (i.e., $\mu_1$, $\mu_3$, and $\mu_4$) respectively. 

Furthermore, the queues at Servers $1$ and $2$ are assumed to have finite buffer. Let $M_{i}$ denote the buffer size of each queue $i \in \mathcal{K}\backslash\left\{6\right\}$. If a queue is full and no task departs at the same time that a new one arrives, the new task is dropped and removed from the system. However, the queue of Server $3$ (where $Q_6$ is located) is assumed to have infinite length of buffer. 
In practice, the buffer in the core network has limited size, which is usually quite large. Our analysis based on the infinite buffer size assumption can capture this scenario with minor modifications. However, we can extract insights on finding the appropriate queue size by performing the analysis assuming infinite queue size.


\section{Performance Analysis}
In this section, we perform the modeling and the performance analysis that allow us to derive the critical  performance metrics. We model the considered queueing system utilizing \ac{DTMC}. Modeling the whole system as one Markov chain can drive in a quite complicated system difficult to be analyzed in terms of closed-form expressions. Thus, in order to simplify the analysis, we decompose the system
into different subsystems. We consider the following four subsystems: 1) $Q_1$ and $Q_2$, 2) $Q_3$ and $Q_4$ 3) $Q_5$, and 4) $Q_6$. The performance metrics for the whole system are approximated with the analytical expressions derived from the subsystems. 
\subsection{Subsystems 1 and 2: Two queues in tandem}
The two queues in tandem $Q_{1}$ and $Q_{2}$ are considered a subsystem. The arrival rate for $Q_{1}$ is: $\lambda_{1}=pa$.
The Markov chain $\left\{(X_{n}\text{, } Y_{n})\right\}$ is described by $P_{i,j:u,k} = \Pr\left\{X_{n+1}=i\text{, } Y_{n+1} = j\text{ }|\text{ } X_{n}=u,Y_{n}=k\right\}$, where $X_n$ and $Y_n$ denote the states (in terms of queue length) of $Q_1$ and $Q_2$ at the $n$-th time slot, respectively, and $i$ and $j$ are referred to as the level $i$ and phase $j$, respectively. The Markov chain is a \ac{QBD} \ac{DTMC} \cite{AppliedDiscrete}. Note that the \ac{QBD} only goes a maximum level up or down, the transition matrix has a block partitioned form:

\begin{small}
	\begin{align}\nonumber
	\mathbf{P}_{1}=
	\left[\begin{array}{ccccc}
	\mathbf{B} & \mathbf{C} & & &  \\
	\mathbf{E} & \mathbf{A}_{1} & \mathbf{A}_{0} & &  \\
	& \mathbf{A}_{2} & \mathbf{A}_{1} & \mathbf{A}_{0}  & \\
	& & \ddots & \ddots &    \\
	&&& \mathbf{A}_{2}  & \mathbf{A}_{0}+\mathbf{A}_{1} 
	\end{array}\right]\text{.}
	\end{align}
\end{small}

For the sake of simplicity, given a probability of an event, denoted by $p$, we denote the probability of its complementary event by $\bar{p}\triangleq 1-p$. To derive the block matrices $\ma{B}, \ma{C}, \ma{E}$ and $\ma{A}_i\text{, } \text{for } i =  0,1,2$,  we first define the following matrices:
First we define the following  matrices 

\begin{scriptsize}
	\begin{align}\nonumber
	\centering
	\mathbf{P}_{1}^{(1)}=
	\left[\begin{array}{cccc}
	1 & 0 & &\\
	\mu_{2} & \bar{\mu}_{2}  & &\\
	& \ddots & \ddots &       \\ 
	& &\mu_{2} &   \bar{\mu}_{2}  
	\end{array}    
	\right]\text{, }
	\mathbf{P}_{1}^{(2)}=
	\left[\begin{array}{ccccccc}
	0 & 1 & 0 & &\\
	0 & \mu_{2} & \bar{\mu}_{2}  & &\\
	0 &  0 & \mu_{2}  &  \bar{\mu}_{2}  &\\
	& & & \ddots  &  \ddots   &  \\
	& & &         &    0       &      1  
	\end{array}
	\right]\text{.}
	\end{align}
\end{scriptsize}

Then, the block matrices of the transition matrix are calculated as 
$$\mathbf{B}= \bar{\lambda}_{1}\mathbf{P}_{1}^{(1)}\text{, } \mathbf{C}=\lambda_{1}\mathbf{P}_{1}^{(1)}\text{, } \mathbf{E}=\bar{\lambda}_{1}\mu_{1}\mathbf{P}_{1}^{(2)}\text{,}$$
$$\mathbf{A}_{0} = \lambda_{1}\bar{\mu}_{1}\mathbf{P}_{1}^{(1)}\text{, } \mathbf{A}_{1}=\bar{\lambda}_{1}\bar{\mu}_{1}\mathbf{P}_{1}^{(1)}+\lambda_{1}\mu_{1}\mathbf{P}_{1}^{(2)}\text{, } \mathbf{A}_{2}=\bar{\lambda}_{1}\mu_{1}\mathbf{P}_{1}^{(2)}\text{.}$$ 
\noindent Following the steps described above, utilizing the properties of a QBD DTMC,  we can construct the transition matrix of Subsystem $1$ for arbitrary finite buffer sizes.

Our goal is to derive the steady state distribution of the Markov chain defined above. We can apply direct methods in order to find the steady state distribution \cite[Chapter 4]{AppliedDiscrete}. Note that there are several efficient algorithms that can be used for this purpose, e.g., logarithmic reduction method. 

We denote the steady state distribution of Subsystem $1$ by a row vector $\boldsymbol{\pi}^{(1)} = \left[\pi_{0,0}^{(1)},\pi_{0,1}^{(1)},\ldots,\pi_{0,M_{2}}^{(1)},\pi_{1,0}^{(1)},\ldots,\pi_{M1,M2}^{(1)}\right]$. We find $\boldsymbol{\pi}^{(1)}$ by solving the following linear system of equations
$\boldsymbol{\pi}^{(1)} \mathbf{P}_{1} =\boldsymbol{\pi}^{(1)}\text{, }  
	\boldsymbol{\pi}^{(1)} \boldsymbol{1} = 1\text{,}
	\label{eqn:steady_state}$
where $\boldsymbol{1}$ denotes the column vector of ones. Hereafter we use $\ve{\pi}^{(n)}$ to denote the steady state distribution vector of the $n$-th subsystem for $n = 1, 2, 3, 4$. 

Furthermore, the arrival rate of $Q_{2}$ depends on the service rate of $Q_{1}$. However, the arrival rate of $Q_{2}$ is equal to $\mu_{1}$ if and only if $Q_{1}$ is non-empty. Therefore, the arrival rate of $Q_{2}$ is
 $\lambda_{2} = \text{Pr}\left\{Q_{1}>0\right\} \mu_{1} = \left(\sum_{j=0}^{M_2} \sum_{i=1}^{M_{1}} \pi_{i,j}^{(1)}\right) \mu_{1}\text{.}
	\label{eq:arr_2}$ Similarly, we can construct the transition matrix $\mathbf{P}_{2}$ and the steady state distribution $\boldsymbol{\pi}^{(2)}$ for the second subsystem consisting of $Q_{3}$ and $Q_{4}$. The arrival rates of $Q_{3}$ and $Q_{4}$ are
$\lambda_{3}  = p (1-\alpha)  \text{ and }$
	$\lambda_{4}   = \text{Pr} \left\{Q_{3}>0\right\} \mu_{3} = \left( \sum\limits_{j=0}^{M_{4}} \sum\limits_{i=1}^{M_{3}} \pi_{i,j}^{(2)} \right) \mu_{3}\text{,}$
respectively.
\subsection{Subsystem 3: $Q_{5}$ with finite buffer}  
We consider $Q_5$ as an independent subsystem. $M_{5}$ is the buffer size of the queue. We first define the arrival rate of $Q_{5}$:
$\lambda_{5} = \text{Pr}\left\{Q_4>0\right\} \mu_{4} = \left(\sum\limits_{i=0}^{M_{3}} \sum\limits_{j=1}^{M_{4}} \pi_{i,j}^{(2)}\right) \mu_{4}\text{.}$

We model the subsystem as one Markov chain whose transition matrix is shown below
\begin{figure}
	\centering
	\includegraphics[scale=0.35]{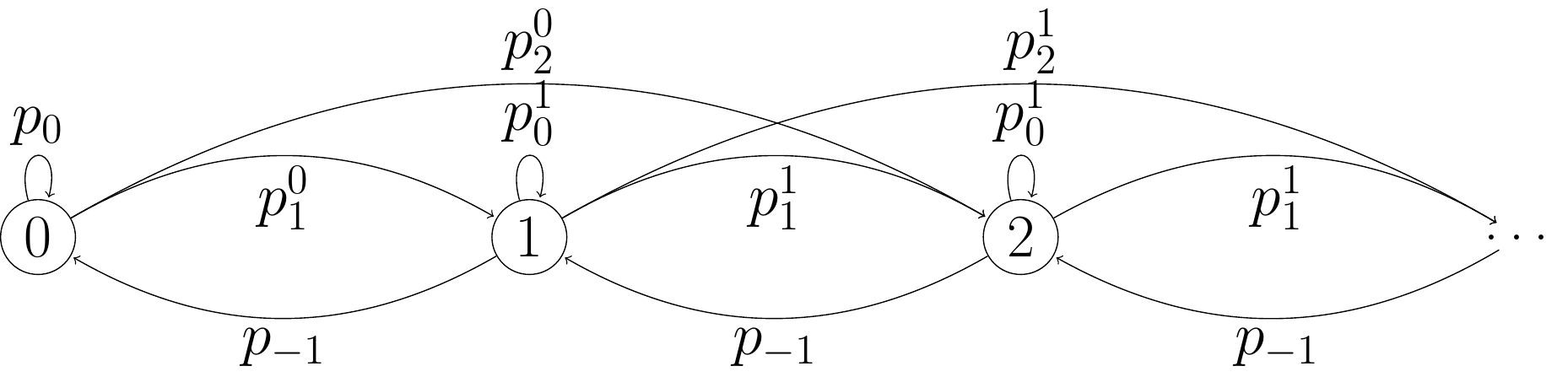}
	\caption{Markov chain for  $Q_5$.}
	\label{Fig: 2BernMarkov}
\end{figure} 
\begin{align}\nonumber
           \mathbf{P}_{3} = 
           \left[\begin{array}{cccc}
     \bar{\lambda}_{5}           & \lambda_{5}                                                       &  &  \\
  \bar{\lambda}_{5}\mu_{5}  & \lambda_{5}\mu_{5}+\bar{\lambda}_{5}\bar{\mu}_{5} & \lambda_{5}\bar{\mu}_{5}  & \\
                                         & \bar{\lambda}_{5}\mu_{5} & \lambda_{5}\mu_{5}+\bar{\lambda}_{5}\bar{\mu}_{5} & \lambda_{5}\bar{\mu}_{5}  \\
                                 & \ddots & \ddots & \ddots    \\
&            &  \bar{\lambda}_{5}\mu_{5} &  \bar{\lambda}_{5}\bar{\mu}_{5} + \lambda_{5}     
\end{array}
\right]\text{.}
\end{align}
We denote the steady state distribution of Subsystem 3 by $\boldsymbol{\pi}^{(3)}= \left[\pi_{0}^{(3)}\text{, }\pi_{1}^{(3)}\text{,}\ldots\text{, } \pi_{M_5}^{(3)}  \right]$.
To derive $\mathbf{\pi}^{(3)}$, we solve the following linear system of equations:
$\boldsymbol{\pi}^{(3)}\mathbf{P} = \boldsymbol{\pi}^{(3)}\text{, } 
\boldsymbol{\pi}^{(3)}\mathbf{1} = 1\text{.}$  Using balance equations, we obtain $\pi^{(3)}_{i} = \frac{\lambda_{5}^i\bar{\mu}_{5}^{(i-1)}}{\bar{\lambda}_{5}^{i}\mu_{5}^{i}}\pi_{0}^{(3)}\text{, for } 1 \leq i  \leq M_{5}\text{,}$
and $\pi^{(3)}_{0} = \left[1+\sum\limits_{i=1}^{M_{5}}\frac{\lambda_{5}^{i}\bar{\mu}_{5}^{i-1}}{\bar{\lambda}_{5}^{i}\mu_{5}^{i}} \right]^{-1}\text{.}$
\vspace{0mm}
\subsection{Subsystem $4$: $Q_6$ with infinite buffer size}
\vspace{0mm}
The arrival rate for $Q_6$ depends on the service rate of $Q_{2}$ and $Q_{5}$, and the probability that the queues are non-empty. Note that the departures from $Q_{2}$ and $Q_{5}$ can be considered independent stochastic processes. The arrival rates that occur due to $Q_{2}$ and $Q_{5}$ are 
	 $\lambda_{6,2}  = \text{Pr}\left\{Q_{2}>0\right\}\mu_{2} \text{ where }
   \lambda_{6,5}  = \text{Pr}\left\{Q_{5}>0\right\} \mu_{5}$,
respectively. The arrival rate of $Q_{6}$ is: $ \lambda_{6} = \lambda_{6,2} + \lambda_{6,5}\text{.}$
We model the system as a Markov chain as shown in Fig. \ref{Fig: 2BernMarkov}, where
\begin{align}\nonumber
	p_{0} & = \bar{\lambda}_{6,2}\bar{\lambda}_{6,5}\text{, }p_{1}^{0}  = \lambda_{6,2}\bar{\lambda}_{6,5} + \lambda_{6,5}\bar{\lambda}_{6,2}\text{, } 	p_{2}^{0}  = \lambda_{6,2}\lambda_{6,5}\label{eqn:Q6_p10}\text{, }\\\nonumber
	p_{0}^{1} & = \bar{\lambda}_{6,5}\bar{\lambda}_{6,2}\bar{\mu}_{6}+\bar{\lambda}_{6,5}\lambda_{6,2}\mu_{6} + \lambda_{6,5}\bar{\lambda}_{6,2}\mu_{6}\text{, } 
	p_{2}^{1}    = \lambda_{6,2}\lambda_{6,5}\bar{\mu}_{6} \\\nonumber
	p_{1}^{1}  & =  \lambda_{6,2}\bar{\lambda}_{6,5}\bar{\mu}_{6} + \bar{\lambda}_{6,2}\lambda_{6,5}\bar{\mu}_{6} + \lambda_{6,2}\lambda_{6,5}\mu_{6}\text{,}\\\nonumber
	p_{2}^{1}  &  = \lambda_{6,2}\lambda_{6,5}\bar{\mu}_{6}\text{, }p_{-1}   = \bar{\lambda}_{6,2}\bar{\lambda}_{6,5}\mu_{6}\text{.}
\end{align}

The transition matrix that describes the Markov chain above is shown below
\begin{align}\nonumber
	\mathbf{P}_{6} = 
	\left[\begin{array}{ccccccccc}
	a_{0}       &      b_{0}     &       0          &         0      &  \cdots  \\
	a_{1}       &      b_{1}     &     b_{0}      &         0      & \cdots   \\
	a_{2}      &      b_{2}     &     b_{1}       &      b_{0}   & \cdots  \\
	0           &     b_{3}      &     b_{2}      &      b_{1}    & \cdots \\ 
	\vdots     &       0          &     b_{3}      &      b_{2}    &    \cdots 
	\end{array}
	\right]\text{,}
\end{align}
where $a_{0}=p_{0}^{0}$, $a_{1}=p_{1}^{0}$, $a_{2}=p_{2}^{0}$, $b_{0}=p_{-1}$, $b_{1}=p_{0}^{1}$, $b_{2}=p_{1}^{1}$, $b_{3}=p_{2}^{1}$. The transition matrix is a lower Hessenberg matrix. We denote the steady state distribution of Subsystem $4$ by $\boldsymbol{\pi}^{(4)}= \left[\pi_{0}^{(4)}\text{, }\pi_{1}^{(4)}\text{, }\ldots\right]$.
The general expression for the equilibrium equations of states is given by the $i$-th term in the following equation: 
 $\pi_{i}^{(4)}=a_{i}\pi_{0}^{(4)}+\sum\limits_{j=1}^{i+1} b_{i-j}\pi_{j}^{(4)}\text{.}$
For the \ac{DTMC} with infinite state space, we apply $z$-transform approach to solve the state equations. The $z$-transforms for the state transition probabilities $a_{i}$ and $b_{i}$ are
$
	A(z) = \sum\limits_{i=0}^{2} a_{i}z^{-i}\text{ and }
	 B(z) = \sum\limits_{i=0}^{3}b_{i}z^{-i}\text{,}
$ respectively. The $z$-transform for the steady state distribution vector $\boldsymbol{\pi}^{(4)}$ is
$
	\Pi(z) = \sum\limits_{i=0}^{\infty}\pi_{i}^{(4)} z^{-i}=\pi_{0}^{(4)}\frac{z^{-1}A(z)-B(z)}{z^{-1}-B(z)}\text{.}
$
The solution for $\pi_{i}^{(4)}$ is given by
\begin{small}
	\begin{align}\nonumber
	\pi_{0}^{(4)}  = \frac{1+B'(1)}{1+B'(1)-A'(1)}\text{, }
	\pi_{i}^{(4)}   =c_{i}  +  \sum\limits_{j=1}^{m} r_{j}(p_{j})^{(i-1)}\text{, } i>0\text{,}
	\end{align}
\end{small}where $r$, $p$, and $c$ are the residues, poles, and direct terms, respectively. Since $Q_6$ has infinite buffer size, we need to characterize the conditions under which the queue is stable.
The Loynes' theorem states: if the arrival and service processes of a queue are strictly jointly stationary and the average arrival rate is less than the average service rate, then the queue is stable.
Therefore, $Q_6$ is stable if and only if the following inequality holds: $\lambda_{6}<\mu_{6}\text{.}$

\subsection{Discussion on the analysis of scaled-up systems}
In this work, we analyze a simple end-to-end system that consists of  three connected servers. We can analyze systems with arbitrary number of servers by decomposing the system into subsystems and analyze each subsystem individually. Then, we use the results of each subsystem in order to derive the analytical expressions for the whole system. A full version of this work can be found in \cite{fountoulakisjournal}.

\section{Key Performance Metrics}
In this section, we provide analytical expressions of the  performance metrics of our interests, i.e.,  system drop rate and average number of tasks of the system by utilizing the results of the previous section. The probabilities to have a dropped task at each time slot for $Q_1-Q_5$ are shown respectively in below

\begin{small}
	\begin{align}\nonumber
	P_{D_{1}} & = \lambda_{1} \bar{\mu}_{1}\sum\limits_{j=0}^{M_2} \pi^{(1)}_{M_{1},j}\text{, }P_{D_{2}}  = \lambda_{2}\bar{\mu}_{2}\sum\limits_{i=1}^{M_1}\pi^{(1)}_{i,M_{2}} \text{, }\\\nonumber
	P_{D_{3}} & = \lambda_{3} \bar{\mu}_{3}\sum\limits_{j=0}^{M_4} \pi^{(2)}_{M_{3},j}\text{, }
	P_{D_{4}}  = \lambda_{4}\bar{\mu}_{4} \sum\limits_{i=1}^{M_{3}}\pi_{i,M_{4}}^{(2)}\text{, }
	P_{D_{5}}  =  \lambda_{5}\bar{\mu}_5 \pi^{(3)}_{M_{5}}\text{,} 
	\end{align}	
\end{small}
where $P_{D_{i}}$ is the probability to have a dropped task of queue $i$.
The average length of each queue is given by
\begin{align}\nonumber
  \bar{Q}_{1}  & = \sum\limits_{i=0}^{M_{1}}\sum\limits_{j=0}^{M_{2}} \pi_{i,j}^{(1)} i\text{, }\bar{Q}_{2} = \sum\limits_{j=0}^{M_{2}}\sum\limits_{i=0}^{M_{1}} \pi_{i,j}^{(1)} j\text{, } 
  \bar{Q}_{3}  = \sum\limits_{i=0}^{M_{3}}\sum\limits_{j=0}^{M_{4}} \pi_{i,j}^{(2)} i\text{, }\\ \nonumber
  \bar{Q}_{4} & = \sum\limits_{j=0}^{M_{4}}\sum\limits_{i=0}^{M_{3}}\pi^{(2)}_{i,j} j\text{, }
  \bar{Q}_{5}  = \sum\limits_{i=0}^{M_5} \pi_{i}^{(3)}i \text{, }
  \bar{Q}_{6} = \sum\limits_{i=0}^{\infty} \pi_{i}^{(4)}i\text{.}  
\end{align}
Therefore, the system drop rate and the average number of task in the system can be described as
\begin{align}\nonumber
      P_{D} = \sum\limits_{i\in\mathcal{K}\backslash \left\{6\right\}}P_{D_{i}} \text{ and }
     \bar{Q} = \sum\limits_{i\in \mathcal{K}} \bar{Q}_{i}\text{, respectively.}
\end{align}


\section{Numerical Results}\label{Sec:Sim}

In this section, we evaluate the accuracy of our derived mathematical model in terms of key performance metrics by comparing the analytical with simulation results. Furthermore, we provide results regarding the system performance under different setups. We developed a MATLAB-based behavioural simulator and each case run for $10^6$ timeslots.
\subsection{Effect of $\mu_{1}$ and $\mu_{2}$ on  the drop rate in systems with small buffers}


\begin{figure}[t!]
 \centering
 \subfloat[Optimal values of $\alpha$. \label{1a}]{%
	\includegraphics[width=0.48\linewidth]{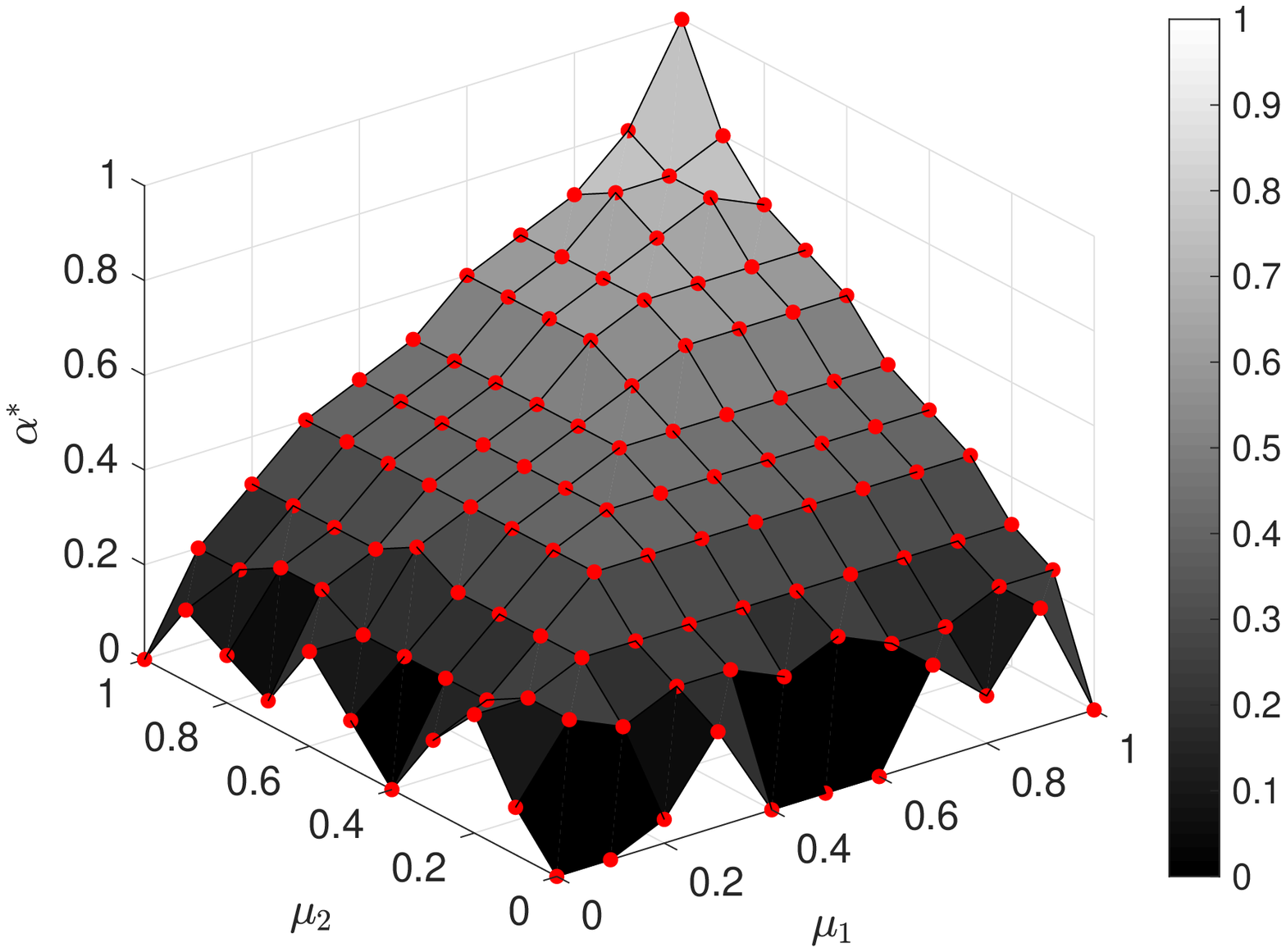}
	\label{Fig: AlphaOptimalDropExp10}}
\hfill
\subfloat[System drop rate. \label{1b}]{%
	\includegraphics[width=0.49\linewidth]{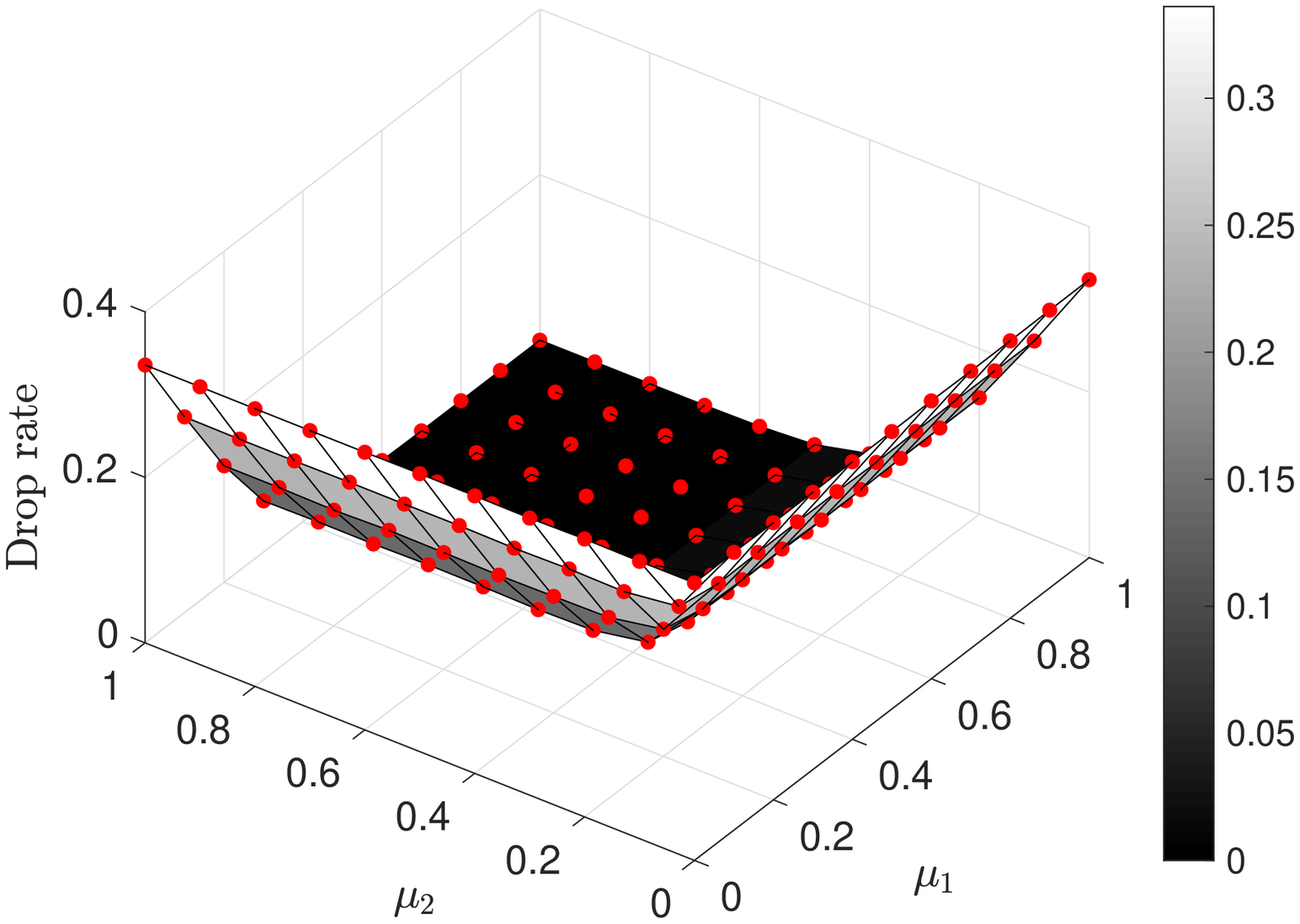}
	\label{Fig: OptimalDropExp10}
	}
	\caption{\small Objective: To minimize the system drop rate. $\mu_{3}=\mu_{4}=\mu_{5}=0.5$, $\mu_{6}=0.9$, $p=0.8$. $M_{i}=10$, for $1\leq i \leq 5$.}
\end{figure}
\begin{figure}[t!]
	\centering
	\subfloat[Optimal values of $\alpha$. \label{1a}]{%
		\includegraphics[width=0.48\linewidth]{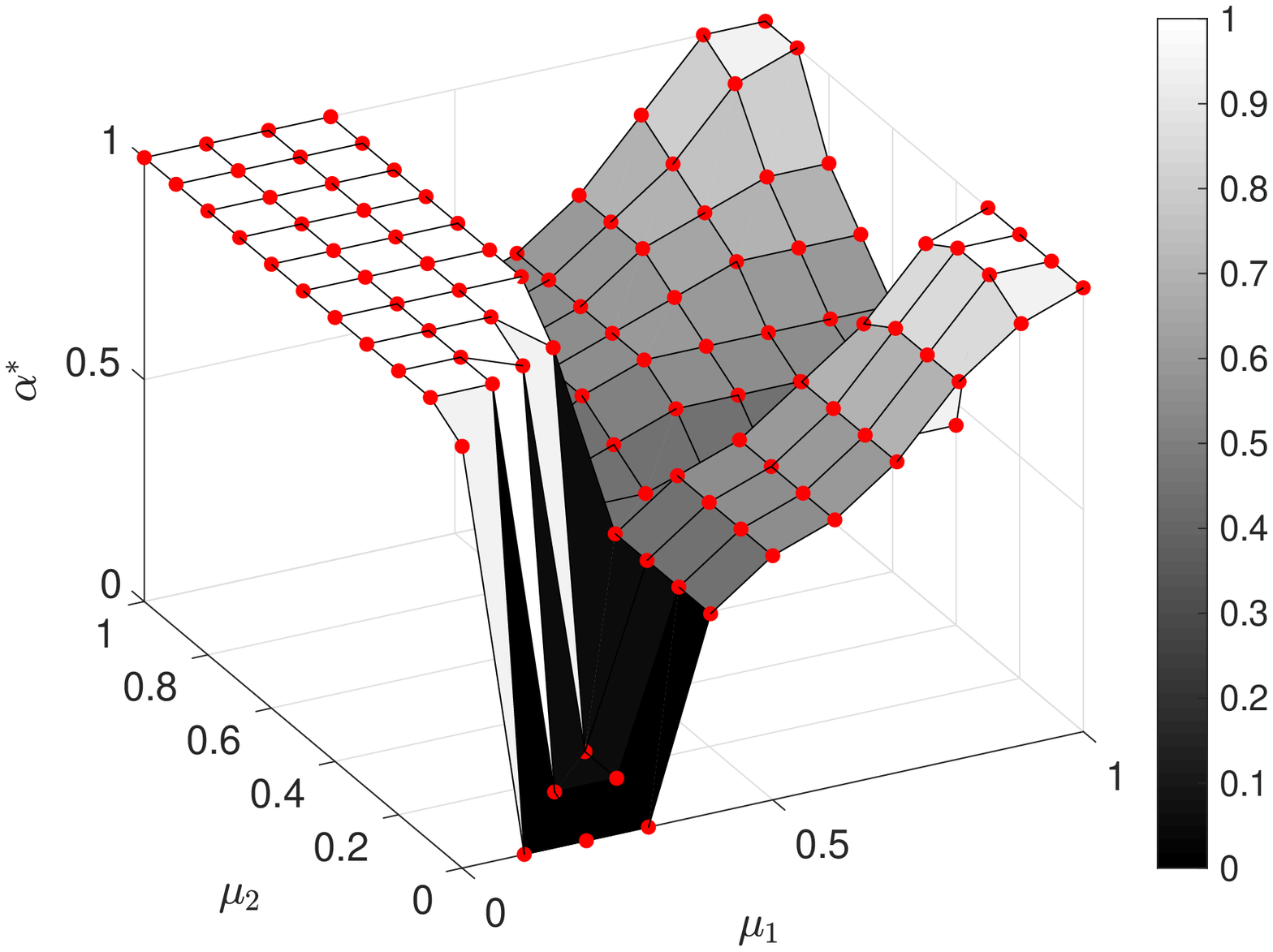}
		\label{Fig: OptimalAlphaTasksExp50}}
	\hfill
	\subfloat[Average number of tasks. \label{1b}]{%
		\includegraphics[width=0.48\linewidth]{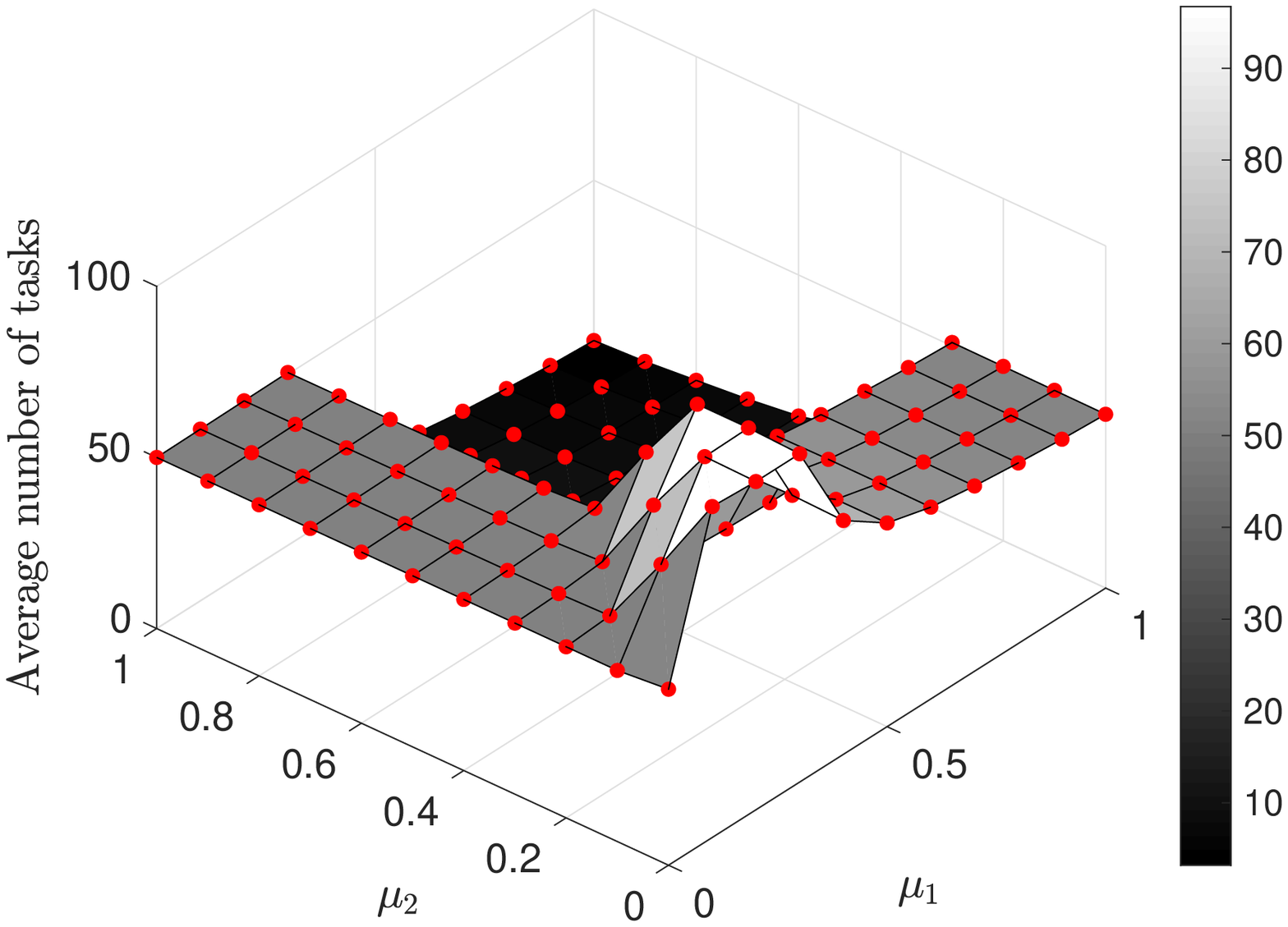}}
	\caption{\small Objective: To minimize the average number of tasks. $\mu_{3}=\mu_{4}=\mu_{5}=0.5$, $\mu_{6}=0.9$, $p=0.8$.  $M_{i}=50$, for
	$1\leq i \leq 5$.}
\end{figure}
In this subsection, we observe the performance of the system in terms of the drop rate when the size of the buffers is small.
In Fig. \ref{Fig: AlphaOptimalDropExp10}, we provide the optimal values of $\alpha$ (probabilistic routing decision) for different values of $\mu_{1}$ and $\mu_{2}$. Note that we obtain the optimal $\alpha$ for each value of $\mu_{1}$ and $\mu_{2}$ by applying brute force. We observe that for small values of $\mu_{1}$ and $\mu_{2}$, the value of $\alpha$ is small (around $0.2$). Therefore, the routing selects to route the traffic flow to the secondary MEC server (Server $2$). Furthermore, it is shown that the value of optimal $\alpha$ is affected by the smaller value between the transmission rate and processing rate. Therefore, the buffer with the smallest transmission or computation capacity becomes the bottleneck for the subsystem. This could be the case, for example, when the connection between the MEC server $1$ and the server in the core network is weak. Fig. \ref{Fig: OptimalDropExp10}  depicts the system drop rate for the corresponding optimal $\alpha$'s.

\subsection{Effect of $\mu_1$ and $\mu_2$ on the number of tasks in systems with large buffers}
In this subsection, we provide results for the performance of the system in terms of average number of tasks. Our objective is to minimize the average number of tasks in the system when the buffer size is large. In Fig. \ref{Fig: OptimalAlphaTasksExp50}, the optimal $\alpha$'s for different values of $\mu_{1}$ and $\mu_{2}$ are shown. We observe that for small values of $\mu_{1}$, the optimal value of $\alpha$ is equal to $1$. The flow controller decides to route the whole traffic to the first server. This decision is optimal in terms of minimizing the average number of tasks in the system, but it increases significantly the system drop rate. The reason is that  a large percentage of the tasks are dropped and but not served. We also observe that the smallest value between $\mu_{1}$ and $\mu_{2}$ operates as bottleneck in the subsystem and subsequently in the whole system.

\subsection{Trade-off between system drop rate and average queue length - simulation vs analytical results}
\begin{figure}[h!]
	\centering
	\includegraphics[scale=0.39]{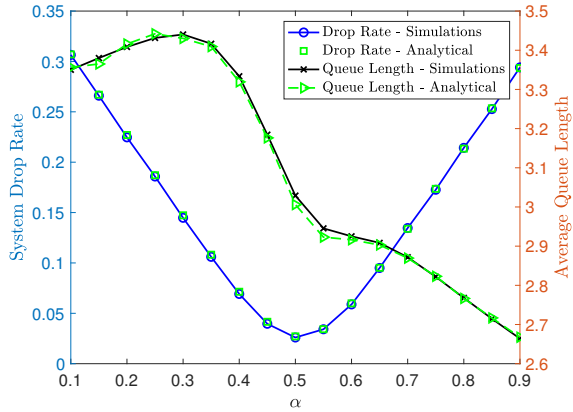}
	\caption{System drop rate and average queue length trade-off. Analytical vs simulation results. $\mu_{i}=0.45$ for $1\leq i \leq 5 $, $\mu_{6}=0.9$. }
	\label{Fig:DropRateVsQLength}
\end{figure}
In this subsection, we provide results that show the trade-off between the system drop rate and average queue length for different routing decisions. In addition, we compare the analytical with simulation results and evaluate the accuracy of our model. In this paper, we show only one scenario due to the space limitation. We observe that $\alpha$ with values around $0.5$ provide the best trade-off. In addition, an interesting result is shown for the case that $\alpha$ takes extreme values, i.e., $0.1$ and $0.9$. Although the system drop rate is almost the same for these two cases, the average queue length is quite different. 
The reason is that  when the first path is selected with higher probability, the traffic flow traverses less number of queues  comparing to the second path. In this case, three of the queues, i.e., $Q_{3}$, $Q_{4}$, $Q_{5}$, are lightly loaded. On the other hand, when the probability the second path to be selected is high, i.e., smaller value of $\alpha_{1}$, the traffic traverses larger number of queue. In this case, more queues are heavily loaded and the average queue length increases.

\subsection{Effect of different buffer capacities and service rates on throughput and delay}
In order to further investigate  the performance of the system, we provide simulation results that show how different setups of the system affect the system throughput and delay. Note that the analytical expressions for the throughput and delay are calculated by using  the results for the system drop rate and average queue length, respectively. We omit the analysis due to space limitation and we  provide the complete analysis in an extended version of this work. However, it is interesting to show some preliminary results. 
\begin{figure}
	\centering
	\includegraphics[scale=0.39]{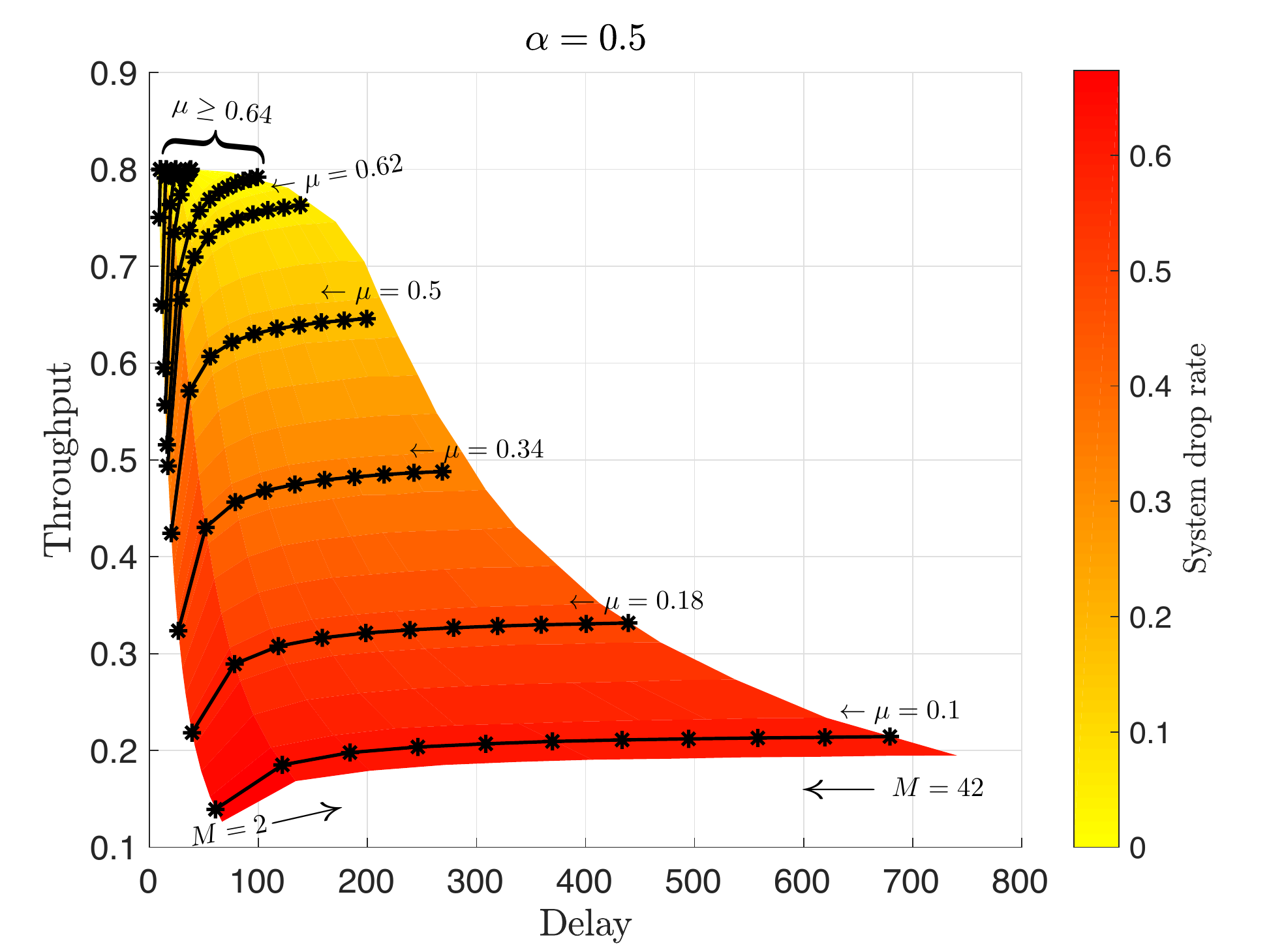}
	\caption{Performance region. $\mu_{i}=\mu$ for  $1\leq i \leq 5$, $M_{i}=M$ for $1\leq i \leq 5$. $p=0.8$, $\mu_{6}=0.9$.}
	\label{Fig. DelayThr}
\end{figure}
In Fig. \ref{Fig. DelayThr}, we provide results for the throughput, delay, and corresponding system drop rate for different values of $\mu$ and $M$. 
We obtain the throughput and delay as following: We fix the service rate $\mu$, and change the capacity of the buffers $M$. Thus, we create each black horizontal line with the stars. The colormap represents the values of system drop rate.

We observe that system performace is significantly affected when we increase the service rates of the buffers. On the other hand, when the service rates are low but the capacities of the buffers are large, the system performance is not improved. From these preliminary results, we observe that it is more important to increase the service rate than the capacity of the buffers.
%

\section{Conclusions \& Future Directions}
In this work, we consider a network topology with two MEC servers, a high-end server at core network, and VNF chains embedded in the servers. We model the network and provide an analytical study on the system performance in terms of system drop rate and average number of the tasks in the system. It is shown, through simulations results, that the approximate model performs well. Numerical results also show useful insights on the design of such systems or resource allocation at each server. Furthermore, we investigate numerically the routing policy that optimizes different objective functions.




\acrodef{NFV}{Network Function Virtualization}
\acrodef{VNF}{Virtual Network Function}
\acrodef{QBD}{Quasi-Birth-and-Death}
\acrodef{MEC}{Multi-access Edge Computing}
\acrodef{DTMC}{Discrete Time Markov Chain}
\acrodef{QBD}{Quasi-Birth-and-Death}
\acrodef{SDN}{Software Defined Network}

\bibliographystyle{IEEEtran}
\bibliography{MyBib1,Bib_QL}
	
\end{document}